\documentclass{webofc}
\usepackage[varg]{txfonts}   
\usepackage{graphicx}
\usepackage{subfig}
\usepackage{amsmath}
\usepackage{dcolumn}
\usepackage{bm}
\usepackage{hyperref}
\usepackage{placeins}
\usepackage{breqn}
\usepackage{multicol}
\newcommand{\fg}[1]{Fig.\hspace{0.7mm}\ref{#1}}

\usepackage{titlesec}
\titlespacing*{\section}{0pt}{0.3\baselineskip}{0.3\baselineskip}
\titlespacing*{\subsection}{0pt}{0.3\baselineskip}{0.2\baselineskip}
\titlespacing*{\abstract}{0pt}{0.5\baselineskip}{0.2\baselineskip}

\begin{document}
%
\title{Heavy flavour energy loss from AdS/CFT: A novel diffusion coefficient}
%
%

\author{\firstname{R.} \lastname{Hambrock}\inst{1}\fnsep\thanks{\email{roberthambrock@gmail.com}} \and
        \firstname{W. A.} \lastname{Horowitz}\inst{1}\fnsep\thanks{\email{wahorowitz@gmail.com}} 
}

\institute{Department of Physics, University of Cape Town, Private Bag X3, Rondebosch 7701, South Africa 
          }

\abstract{%
	Two AdS/CFT based energy loss models are used to compute the suppression and azimuthal correlations of heavy quarks in heavy ion collisions.
	The model with a velocity independent diffusion coefficient is in good agreement with B and D meson data up to high $p_T$.
	The partonic azimuthal correlations we calculate exhibit an order of magnitude difference in low momentum correlations to pQCD calculations \cite{arXiv:1305.3823}. We thus propose heavy flavour momentum correlations as a distinguishing observable of weakly-  and strongly-coupled energy loss mechanisms.
	The partonic azimuthal correlations we calculate exhibit an order of magnitude difference in low momentum correlations to pQCD calculations \cite{arXiv:1305.3823}. We thus propose heavy flavour momentum correlations as a distinguishing observable of weakly-  and strongly-coupled energy loss mechanisms.
}
\maketitle
\section{Introduction}
\label{sec:Introduction}
The quark gluon plasma is of great interest since it represents our first case study of the emergent physics of the non-abelian gauge theory QCD. A key step in understanding this state of matter is identifying its relevant coupling strength. The perturbative techniques of QCD are only adequate in a weakly coupled plasma, with calculations for strongly coupled plasmas constrained to methods like AdS/CFT-based approaches or Resonance Scattering \cite{arXiv:1101.0618}. 

Both strong- and weak-coupling show qualitative agreement with $R_{AA}^D$ \cite{arXiv:1210.8330}, suggesting that they have attained sufficient maturity to compare them with more differential observables. We will argue that the momentum correlations of bottom quarks constitute a promising candidate as a differentiator between weakly and strongly coupled plasmas.

We will use two different AdS/CFT based energy loss models, one \cite{arXiv:1501.04693} with a velocity dependent diffusion coefficient and the other \cite{arXiv:1605.09285} with a diffusion coefficient independent of the heavy quark's velocity. Futhermore, we will probe the spectrum of their possible predictions with two plausible \cite{hep-th/0612143} ’t Hooft coupling constants ($\lambda_1 = 5.5$ and $\lambda_2=12\pi\alpha_s\approx11.3$ with $\alpha_s=0.3$) where for the former, temperature and the Yang-Mills coupling are equated, while for the latter constant, energy density and the coupling are equated.

We will compare these with the pQCD based azimuthal correlations calculated in \cite{arXiv:1305.3823}. These will provide a secondary indicator for the momentum correlations.

Finally, we will compute the suppression of B and D mesons and compare with measurements from CMS \cite{Sirunyan:2017oug} and ALICE \cite{ALI-PREL-128542} respectively.

\section{\label{sec:EnergyLossModel}Energy Loss Model and Suppression}
\subsection{\label{subsec:Overview}Overview}
Subsequent to initializing the momenta of heavy quark pairs either to leading order with FONLL \cite{arXiv:1205.6344} or to next-to-leading order with aMC@NLO \cite{arXiv:1405.0301} using Herwig++ \cite{arXiv:0803.0883} for the showering and hadronization, the production points of the heavy quarks are weigthed by the Glauber binary distribution \cite{arXiv:1501.04693}. The particles are propagated through the plasma via the energy loss mechanism described in 2.2 until the temperature in their local fluid cell drops below the Tc threshold and hadronization is presumed to occur or 8.6fm had passed, being the maximum time of the VISHNU background \cite{arXiv:1105.3226}. 

\subsection{\label{subsec:LangevinEnergyLoss}Langevin Energy Loss}
The stochastic equation of motion for a heavy quark in the fluid's rest frame is \cite{hep-ph/0412346}
\begin{equation}
  \frac{dp_i}{dt}=-\mu p_i + F_i^L + F_i^T \label{eqn:stochastic_eom}
\end{equation}
where $F_i^L$ and $F_i^T$ are longitudinal and transverse momentum kicks with respect to the quark's direction of propagation and with $\mu$, the drag loss coefficient, being given by $\mu=\pi \sqrt{\lambda} T^2/2M_Q \label{eqn:drag}$ \cite{hep-th/0605158}
where $M_Q$ is the mass of a heavy quark in a plasma of temperature $T$ with 't Hooft coupling constant $\lambda$. The correlations of momentum kicks are given by
\begin{multicols}{2}
	\noindent
	\begin{equation}
		\langle F_i^T(t_1)F_j^T(t_2) \rangle = \kappa_T(\delta{ij}-\frac{\vec{p}_i\vec{p}_j}{|p|^2})g(t_2-t_1) \label{eqn:F_T}
	\end{equation}
	\begin{equation}
		\langle F_i^L(t_1)F_j^L(t_2) \rangle = \kappa_L\frac{p_ip_j}{|p|^2}g(t_2-t_1) \label{eqn:F_L}
	\end{equation}
\end{multicols}
where g is only known numerically \cite{arXiv:1501.04693} and with \\
\begin{minipage}{.5\linewidth}
\begin{equation}
		\kappa_T=\pi\sqrt{\lambda}T^3\gamma^{1/2}
\end{equation}
\end{minipage}
\begin{minipage}{.5\linewidth}
\begin{equation}
		\kappa_L=\gamma^2\kappa_T=\pi\sqrt{\lambda}T^3\gamma^{5/2}
\end{equation}
\end{minipage} 
\vskip.5\baselineskip
It should be noted that this construction does not obey the fluctuation-dissipation theorem \cite{arXiv:1501.04693}.
The computations based on this model will be labeled \textit{D(p)}.

\subsection{\label{subsec:LangevinEnergyLossMoerman}Development on energy loss model}

The problem with the energy loss mechanism described in \ref{subsec:LangevinEnergyLoss} is that since the longitudinal momentum fluctuations grow as $\gamma^{\frac{5}{2}}$, our setup breaks down for high momenta, where in a perturbative QCD setting, Brehmstrahlung would restrict the momentum growth of the quark. 
Via a novel calculation presented in \cite{arXiv:1605.09285, Hambrock:2017zpa, arXiv:1703.05845}, we instead consider a stationary string in $AdS_d$ hanging into a black hole horizon and calculate $s^2(t,a,d)$ of the free endpoint.
For the the $d=3$ result, the average squared distance travelled can be determined analytically for small string lengths, which is identical to the asymptotically late time behavior of a string with arbitrary initial length. We thus find the asymptotically late time behavior of a string in d dimensions by
\begin{eqnarray}
  s^2(t \gg \beta, a, d) = s_{\mbox{small}}^2(t \gg \beta, a, d) = (\frac{d-1}{2})^2s_{\mbox{small}}^2(t \gg \beta, a, d=3) =
  \frac{(d-1)^2}{8\pi\sqrt{\lambda}}\beta(1-\frac{a}{2})
\end{eqnarray}
where $\beta=T^{-1}$ and $a$ parametrizes between a heavy quark for $a=0$ and a light quark for $a=1$.
At late times, the motion is diffusive, thus we can extract the diffusion coefficient
\begin{equation}
  D(a, d) \sim \frac{1}{2}s^2(t \gg \beta, a, d)
\end{equation}
which in $AdS_5$ for a heavy quark reads $2\beta/\pi\sqrt{\lambda}$. From this, we obtain 

\noindent\begin{minipage}{.5\linewidth}
\begin{equation}
		\kappa_T=2T^2/D=\pi\sqrt{\lambda}T^2/\beta=\pi\sqrt{\lambda}T^3  
\end{equation}
\end{minipage}%
\begin{minipage}{.5\linewidth}
\begin{equation}
		\hat{q}=\langle p_\perp(t)^2 \rangle{\lambda} \approx \kappa_Tt/\lambda = (\pi T^3 t)/\sqrt{\lambda}
\end{equation}
\end{minipage}
\vskip.5\baselineskip
Requiring these fluctuations to obey the fluctuation-dissipation theorem, we attain $\mu = \pi \sqrt{\lambda} T^2 / 2E$. The computations based on this model will be labeled \textit{D=const}.

  \section{Azimuthal correlations and $R_{AA}$}
  \label{sec:Azimuthal correlations}
  In \cite{arXiv:1305.3823}, at leading order, the weak coupling based computations exhibited very efficient broadening of initial azimuthal correlations for low $p_T$ $b\bar{b}$ pairs ($[4-10]\text{ GeV}$), which were washed out once NLO production processes were taken into consideration. 

  Both for mid- and high-$p_T$ ($[4-10]\text{ GeV}$ and $[10-20]\text{ GeV}$ respectively), the initial correlations survive to a large degree, both at leading order and at next-to-leading order, suggesting that they may still be observable in an experimental context.
  
  We compare our strong coupling azimuthal correlations to the weak coupling ones in \fg{fig:AzimuthalCorrelations}. For $[10-20]\text{ GeV}$, our correlations are significant more peaked at their initial back-to-back correspondence. At $[4-10]\text{ GeV}$, this observation still holds for the upper bound of our parameters with $\lambda_1=5.5$, while the $\lambda_2=11.3$ bounded result is of similar magnitude but looser angular correlation than either the collisional or the collisional + Bremsstrahlung based results. In the $[1-4]\text{ GeV}$ range, the azimuthal correlations are almost entirely washed out for $\lambda_2=11.3$, while for $\lambda_1=5.5$, they are broadened with similiar efficiency to the weak coupling results.

  \begin{figure}
	   \subfloat{\includegraphics[width=.35\textwidth]{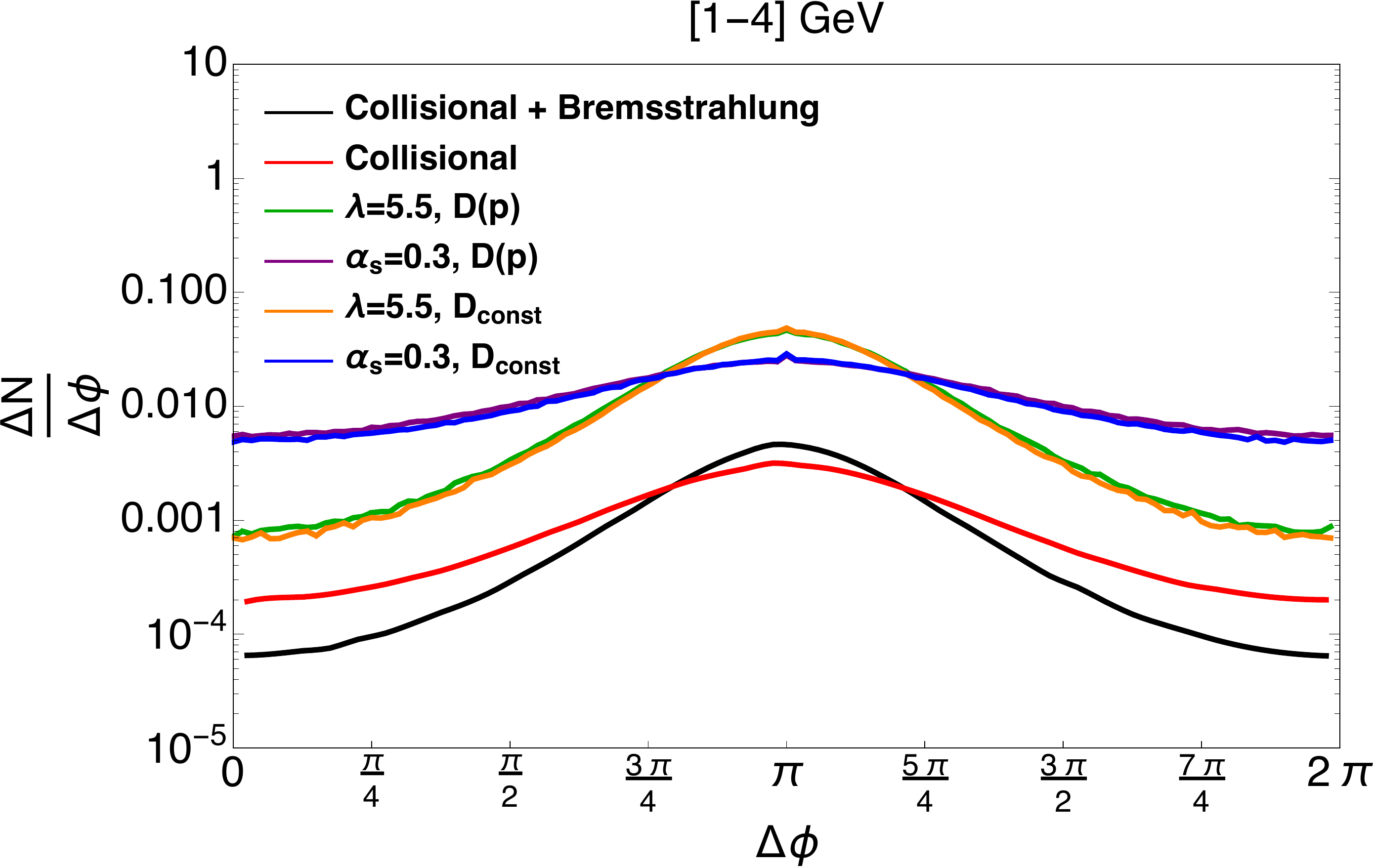}}
	   \subfloat{\includegraphics[width=.35\textwidth]{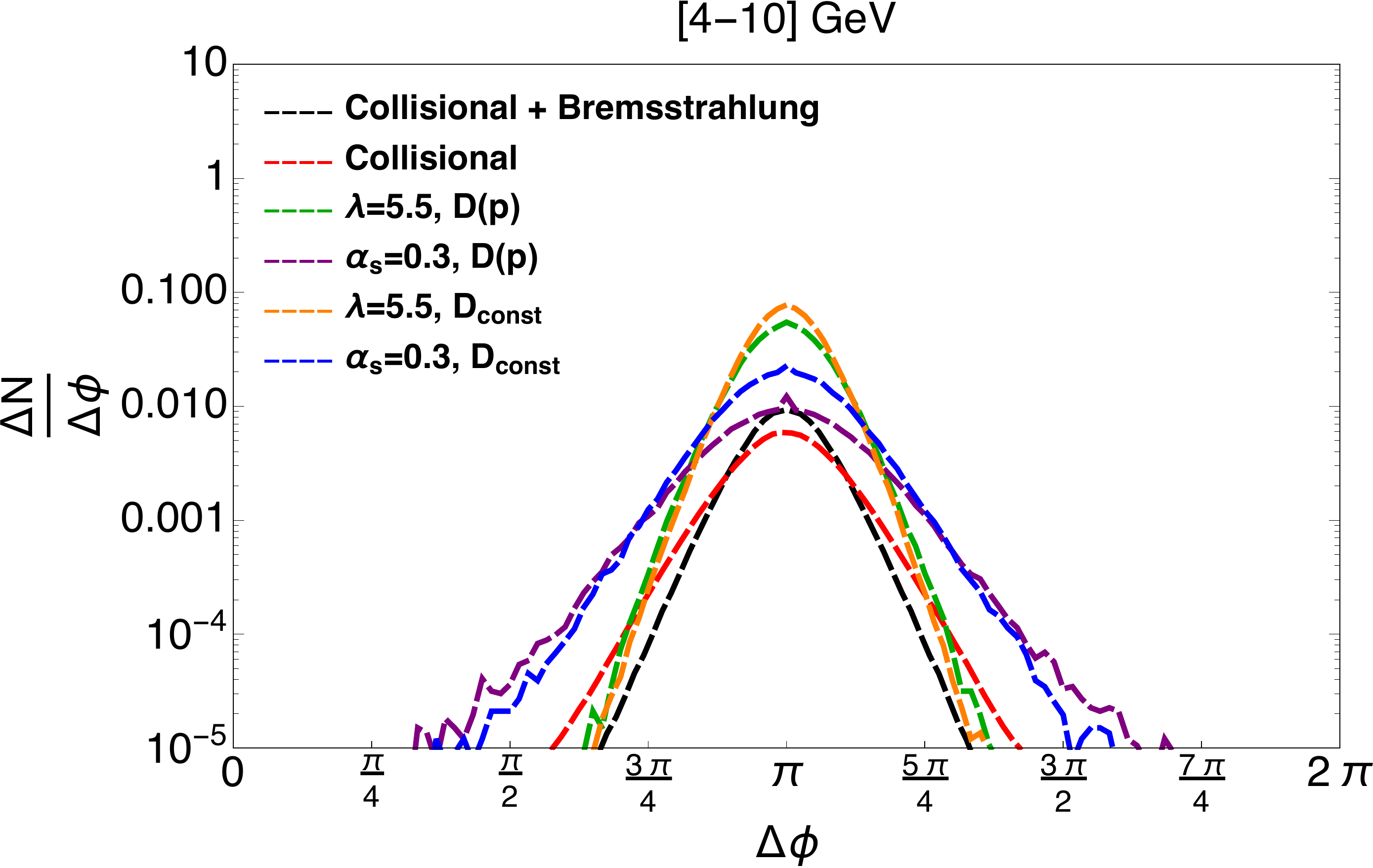}}
	   \subfloat{\includegraphics[width=.35\textwidth]{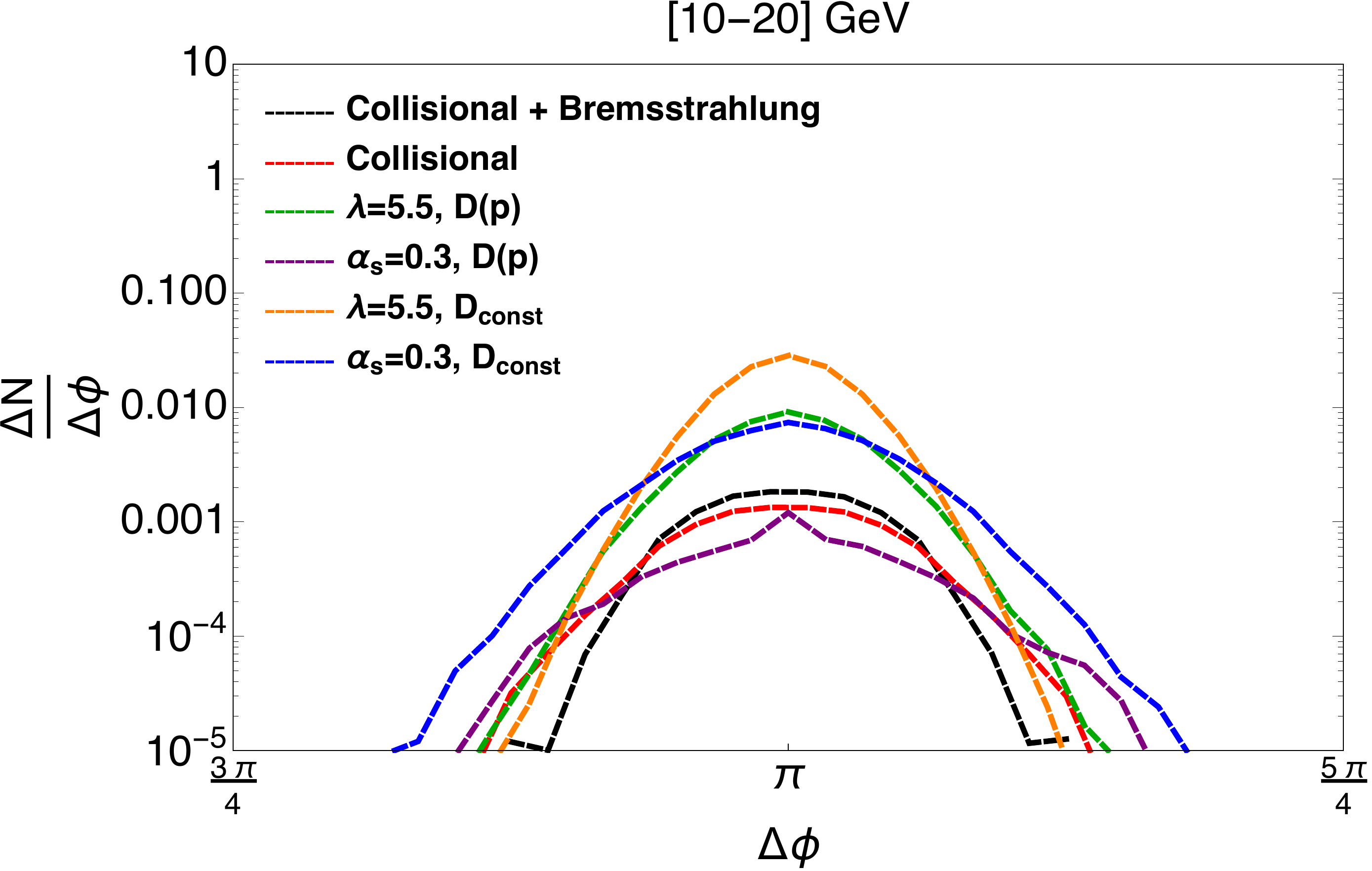}}
    \caption{Bottom quark $\frac{dN}{d\phi}$ correlations for the specified classes.}
    \label{fig:AzimuthalCorrelations}
  \end{figure}


  Having compared our strong coupling results with pQCD calculations at leading order, we now turn to a more realistic simulation that we can compare with data. The results shown in \fg{fig:RAA} are from the next-to-leading order procedure described in \ref{subsec:Overview}.

  In \fg{fig:RAA}, we compare our calculations for the suppression of B mesons with measurements from CMS. At low $p_T$, both \textit{D(p)} and \textit{D=const} models are consistent with data. At high $p_T$, the agreement is inconclusive.
  
  Conversely, for the calculation of averaged $D^0$, $D^+$, and $D^-$, the \textit{D(p)} model diverges from ALICE data \cite{ALI-PREL-128542}. This divergence is due to the momentum fluctuations going as $\gamma^{\frac{5}{2}}$. The \textit{D=const} model remains consistent with data even for high-$p_T$.

  
  \begin{figure}[ht]
	  \subfloat
	  {\includegraphics[height=160px]{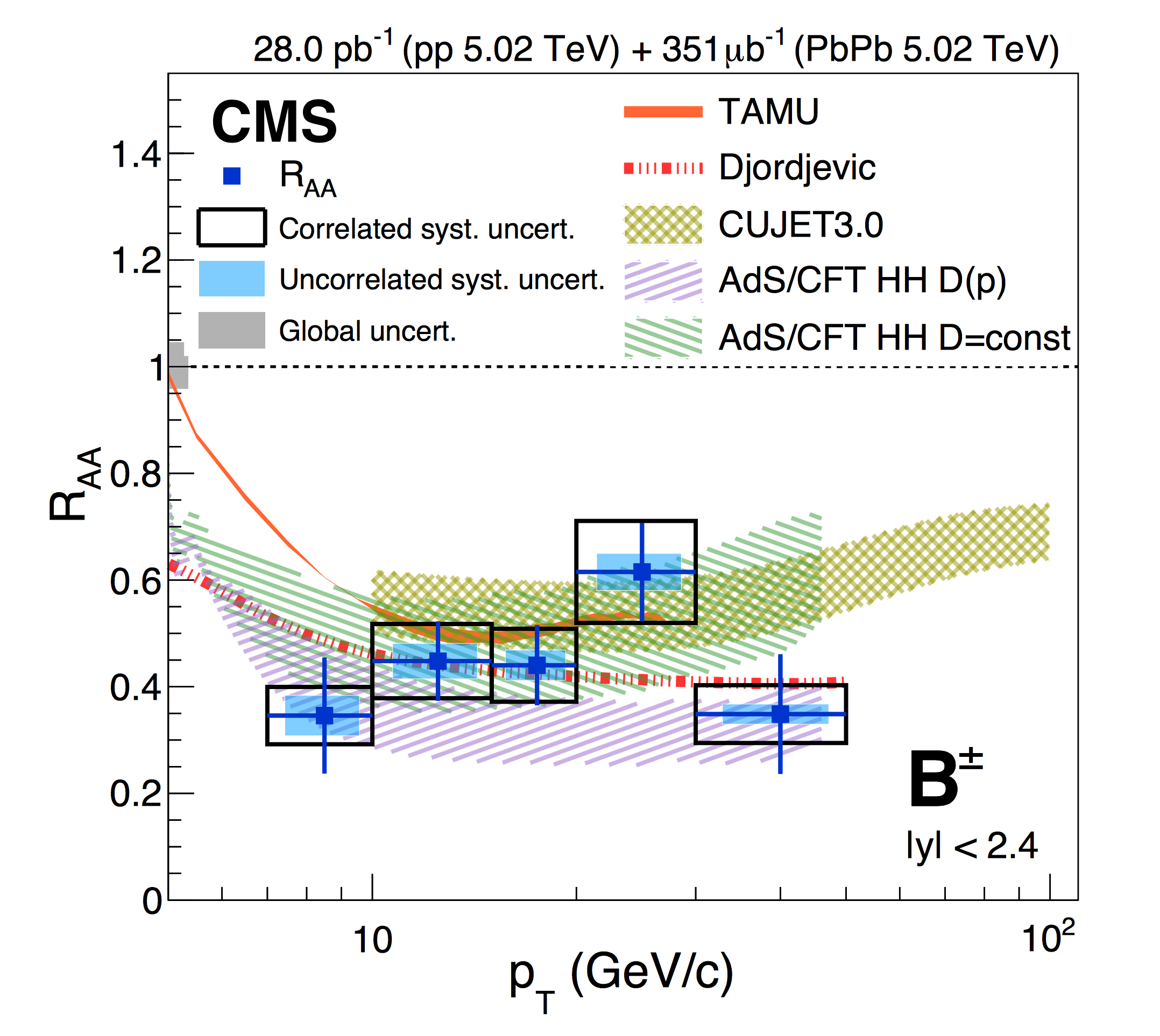}}
	  \subfloat{\includegraphics[height=150px]{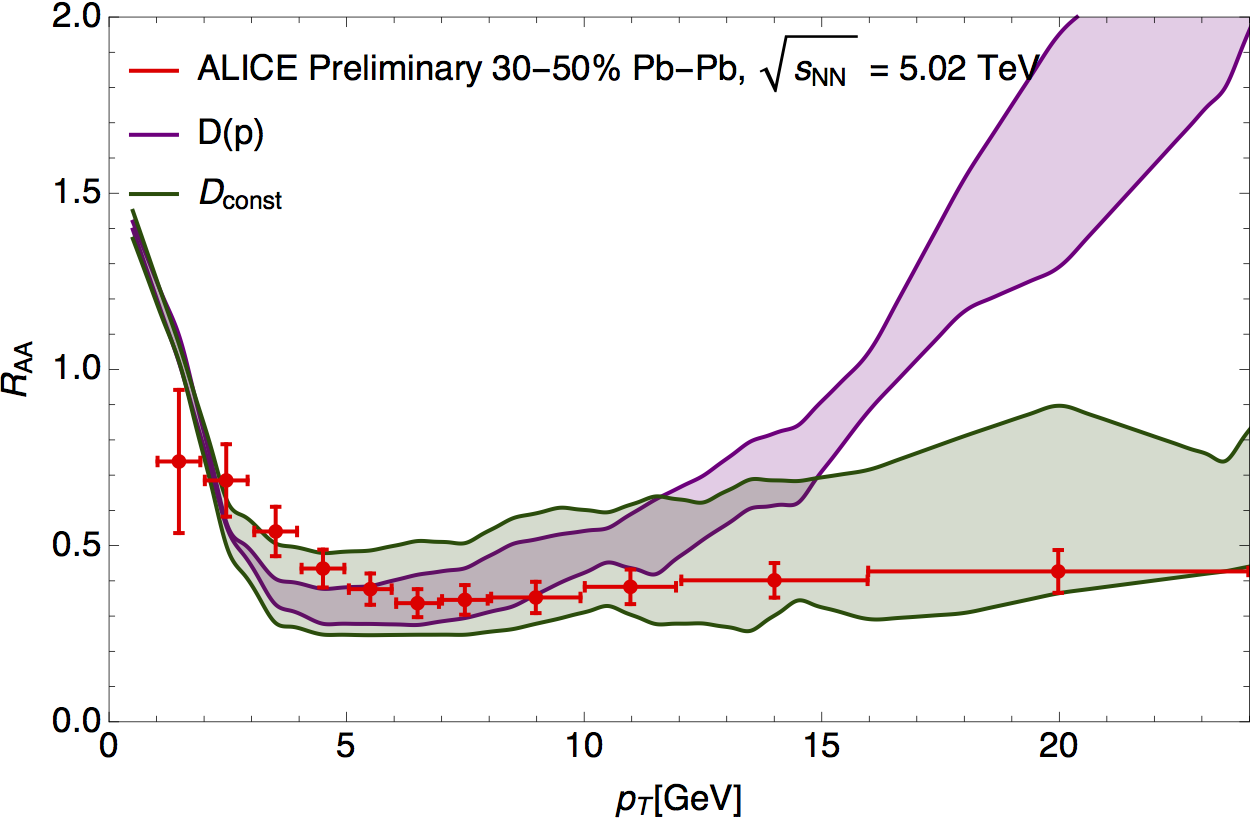}
	  }
	  \caption{(Left) Comparison with $R_{AA}^B$ data from CMS \cite{Sirunyan:2017oug} with $\sqrt{s_{NN}} = 5.02\text{ TeV}$, $\lvert\mbox{y}\rvert < 2.4$. (Right) Comparison with $R_{AA}^D$ data from ALICE \cite{ALI-PREL-128542} with $\sqrt{s_{NN}} = 5.02\text{ TeV}$, $\lvert\mbox{y}\rvert < 0.5$.}
		    \label{fig:RAA}
		    \end{figure}

\section{Conclusion \& Outlook}
We have compared the azimuthal correlations predicted by pQCD and AdS/CFT based computations and found that, while the azimuthal correlations are qualitatively similar, the momentum correlations tell a different tale.
In particular, the surprise of our findings is the large dissimilarity in low momentum correlations of the pQCD and AdS/CFT based simulations; see \fg{fig:AzimuthalCorrelations} (left). Thus, bottom quark momentum correlations present an opportunity to distinguish between the energy loss mechanisms of the two frameworks. 

Although stronger momentum fluctuations in the weakly coupled plasma have been identified as a plausible explanation for this disparity, it can only be verified by computing initial correlations for the weak coupling based correlations as well. 
Furthermore, whether this order of magnitude difference in predictions for low $p_T$ correlations of bottom quarks exposes weaknesses in either or both of the frameworks cannot be declared until experimental data of bottom quark momentum correlations emerge. 

While the agreement with CMS data for B meson suppression is comparable between the \textit{D(p)} and \textit{D=const} models, the comparison with ALICE data for D mesons shows the limited validity range of the \textit{D(p)} model, whereas the \textit{D=const} model remains consistent with data even for high-$p_T$. More fundamentally, for the \textit{D(p)} model, the AdS/CFT picture naturally breaks down at $p_T\sim100\text{ GeV}$ \cite{arXiv:1501.04693}. For the \textit{D=const} model, there is no such natural breakdown. Only for asymptotically large $p_T$ \emph{and} $T$ is one guaranteed that the physics is perturbative.

\section{Acknowledgements}

The authors wish to thank the South African National Research Foundation and SA-CERN for their generous financial support.

\bibliography{MomentumCorrelationsPaper}

\end{document}